# COMPUTATIONAL FLUID DYNAMIC SIMULATIONS IN A MODEL OF A CAROTID BIFURCATION UNDER STEADY FLOW CONDITIONS

Nurullah Arslan[1,*], Kazım Beşirli[2], Gürcan Cömert[3], Omer F. Beyca[4]

[1,4]Fatih University, Industrial Engineering Department, B. Çekmece, Istanbul, Turkey, 34500
[2]Cerrahpaşa Medical School, Surgery Department, Istanbul, Turkey, 34303
[3]University of South Caroline, Civil and Environmental Engineering, Columbia, SC 29208, USA

**Abstract:**
Strokes are still one of the leading causes for death after heart diseases and cancer in all over the world. Most strokes happen because an artery that carries blood uphill from the heart to the head gets clogged. Most of the time, as with heart attacks, the problem is atherosclerosis, hardening of the arteries, calcified buildup of fatty deposits on the vessel wall. The primary troublemaker is the carotid artery, one on each side of the neck, the main thoroughfare for blood to the brain. Only within the last 25 years, though, have researchers been able to put their finger on why the carotid is especially susceptible to atherosclerosis. In this study, the fluid dynamic simulations were done in a carotid bifurcation under the steady flow conditions computationally. In vivo geometry and boundary conditions were obtained from a diseased who has both sides of stenosis located in his carotid artery patients. The location of critical flow fields such as low wall shear stress (WSS), stagnation regions and separation regions were detected. Low WSS was found at the downstream of the bifurcation.

**Introduction:**

The formation of atherosclerosis has been reported due to the low or high shear regions. These regions develop at the downstream of the carotid bifurcation which is located in bifurcation part of the carotid arteries in humans. The low and high shear regions correlation with the atherosclerosis region were discussed (Fry 1968, Caro 1971) and stated the close correlation to the formation of lesions (Caro et al., 1971, Ku et al., 1985, Friedman 1989, Korner et al., 1999). Arterial bifurcation studies have shown that the stress level in the bifurcating areas is several times higher than that in a straight segment. The effect of the arterial wall tension in the formation of atherosclerosis was proposed by Willis (1954) and also by Glagov et al. (1961), who suggested that atherosclerotic plaques occurred more frequently at wall locations where wall tension would be expected to be elevated. More researchers such as (DeBakey et al., 1985, Thubrikar at al., 1995) reported that increased tensile stress predisposes tissue to atheroschelorosis. Nevertheless, most of theses studies have been confined to either the arterial wall itself ignoring the pulsatile flow of blood, or the blood flow domain alone assuming non-distensible walls. There only a few numerical studies that have treated both pulsatile flows and compliant walls (Lou and Yang 1993, Perktold and Rappitsch, 1995, Cebral et al., 2001)

## 2. Materials and Methods

A male patient who has two sided atherosclerosis was the one who will be prepared for the surgery, Figure 1. His age was 50 years old. He was smoker and normotensive. After color Doppler measurement, he was taken to the surgery. Imaging of the left carotid artery was carried out with color Doppler ultrasound (taken at Cerrahpasa Medical School, Istanbul).

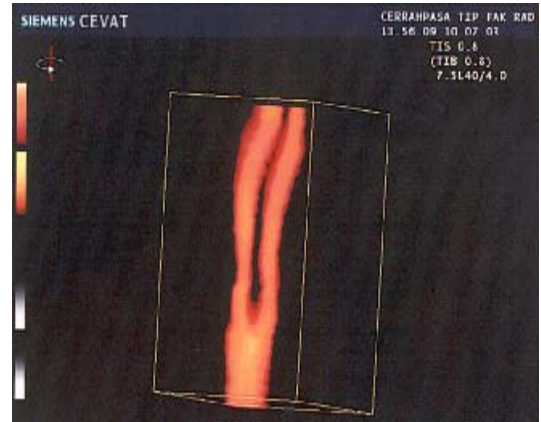

**Figure 1**. The geometry of the stenosed carotid artery taken by Color Doppler Ultrasound Imaging technique

After imaging the flow in color Doppler scan mode, a static Doppler sample volume was deposited in the center inlet of the carotid and the bifurcating vessels, spanning approximately one third of the vessel diameter. A marker was aligned with the vessel axis, and the angle between the Doppler line of sight and the vessel axis was computed. Since the Doppler scan plane was arranged to be that containing the axes of the vein and the graft at the junction, this measured angle was the true angle between the Doppler line of sight and the vessel axis. It was therefore possible to compute true axial velocities. The velocity trace for the graft showed much higher velocities than in normal arteries, with a low pulsatility index since the sonograph trace remained well above zero throughout the cycle. For the patient shown in Figure 1, the instantaneous mean of the sonograph trace at the inlet was estimated to be 0.68 m/s average, maximum 1m/s and 0.2 m/s minumum.. Measurements were also made in the arterial branch site. The velocity value was 2 m/s at the most stenosed location and 1.3 m/s at downstream of the stenosis, Figure 2-A-B-C. Spectral broadening was detected at the most stenosed location. Carotid model was created on the basis of images supplied by the cardiac surgeons representing the typical alternative designs, Figure 1. 3-D model of typical carotid arteries for a diseased patients was utilized using ACIS-based solid modeller Gambit (Fluent Inc., Lebanon, NH, USA) where the 3-D model and mesh were also created, Figure 3. The diameter of the entrance artery was taken as five millimeters in diameters from the color Doppler image. The model was meshed with an automatic meshing algorithm using tetrahedral elements.

*Corresponding author.
E-mail: narslan@fatih.edu.tr or narslan@mit.edu

Reynolds number for the carotid inlet is difficult to estimate, since the shape of the velocity profile is unknown and the sample volume did not cover the whole vessel lumen. Based on the velocity measured in the sample volume (V), and assuming normal blood viscosity ($\mu = 3.5$ mPa s), the above measurements lead to a systolic peak Reynolds number of 1020 and a diastolic minimum of 300 ($Re = \rho VD/\mu$, where $\rho$ = blood density, 1050 kg/m$^3$). The entrance velocities were selected as 0.68 m/s at average, and 0.2 m/s at diastolic flow, respectively. The commercial finite volume code Fluent, Fluent Inc., Lebanon, NH, USA was used in this study. A segregated solver was selected and convergence criteria was selected as $10^{-8}$. The model was meshed with an automatic meshing algorithm using tetrahedral elements. Simulations were performed on a Pentium (R) 4 CPU 2,800 MHz with 1,000 MB of RAM.

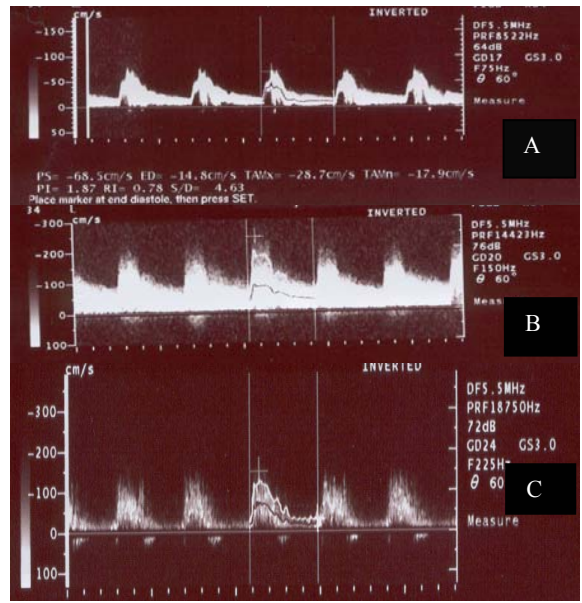

**Figure 2.** Velocity profiles inside the different regions of the carotid artery taken by color Doppler Ultrasound. A.Inlet B. Stenosis C. Downstream of the stenosis

Two assumptions are here implicit: 1) that the velocity profile is flat, and 2) that the patient has normal haematocrit. The true profile is something between flat and parabolic; if one assumes a relationship between spatially averaged and centerline velocity midway between these extremes, the Reynolds number is reduced by 25%. If viscosity is reduced by 20% as suggested in the introduction, the Reynolds number is increased by 25% (1/0.8 = 1.25). The true *in vivo* Reynolds number therefore remains unknown however, the range appears to span the laminar in a circular pipe ($Re_{critical} \approx 2300$).

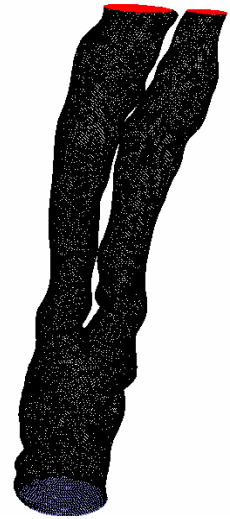

**Figure 3**. Mesh formation for diseased carotid artery.

**Results:**

The fluid flow characterization of the flow inside the carotid bifurcation was analyzed computationally. Figure 4 shows the velocity vectors inside the diseased carotid artery together with the internal and external arteries. SI units were used. The flow enters the model with a flat velocity profile. The peak Reynolds number was 1020 at the inlet. The flow goes through internal and external carotid arteries and accelarated through the internal and external carotid arteries due to the reduction in cross-sectional area. The flow separation regions downstream of the branching point was seen by retrograde vectors. The artery stenosis on both left and right carotid arteries were seen in this specific patients. The velocities inside the stenosis region is increased similar to the jet flow. Depending on the internal geometry of the diseased artery, the circulation regions were seen not only at the inner surface of the branches but also on the external surface of the branches. The maximum velocity was found as 1.7 m/s at the narrowest region of the flow inside the internal carotid artery. The endothelial cells inside the arteries are very sensitive to the high and low shear stress regions. The regions found here would cause the increase of the intimal hyperplasia if there was no surgery done which will also cause the increase of the thickness of the diseased region more. This will cause the sudden ichemia for the patient. which may result the patient to get killed or to get his brain affected. High velocity flow regions were found in the diseased artery due to the stenosis formation. The cross-sectional velocity profiles throughout the complete model and at some specific points such as the carotid branching points and inside the internal and external branching points were shown in Figure 5-A-B. The plots clearly shows the nature of secondary flow inside these regions. The shear stress distribution on the wall is illustrated in Figure 6. The highest WSS was seen at downstream of the carotid branching. The lower WSS region was found where the fluid circulation occurs on both sides of the branches. The structure of the internal part of the carotid was higly complex after stenosis formation. The pressure difference between the inlet and outlets of the model was around 500 Pa shown in Figure 7. This is also due to the resistance of the geometry seen in side the artery inlet and branches. The difference in a healthy model will be less than the one seen in the diseased patient.

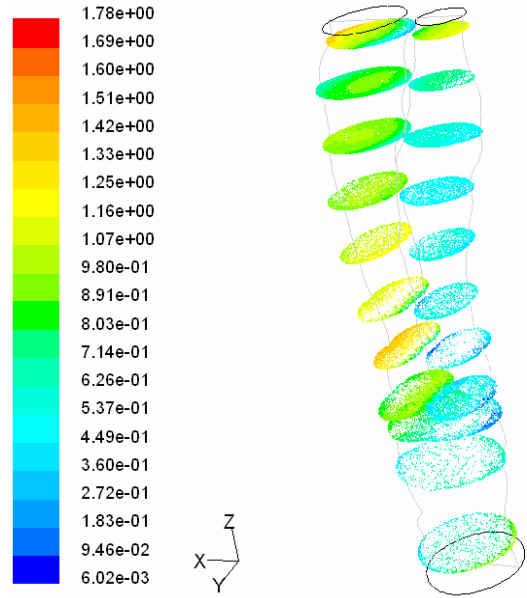

**Figure 5-A**. Cross sectional velocity profiles for complete model b.at the branching point and inside the branches (unit, m/s)

=

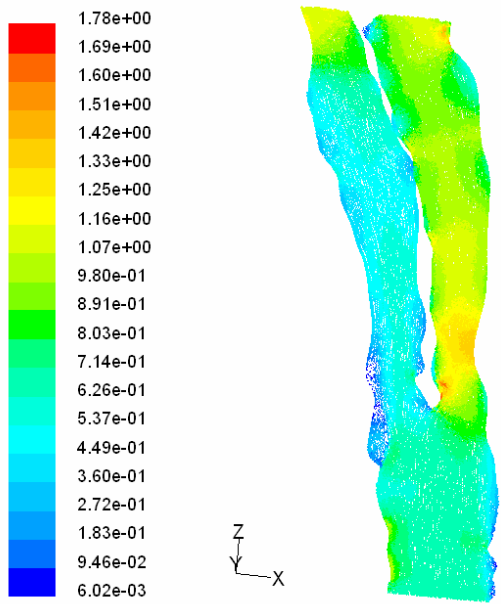

**Figure 4**. The velocity profile inside the diseased carotid artery, in the midplane (unit, m/s)

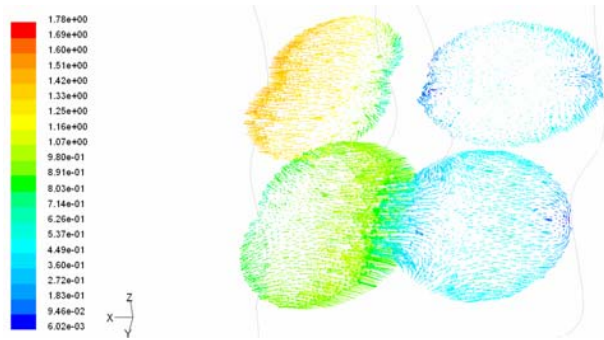

**Figure 5-B**. Cross sectional velocity profiles at the branching point and inside the branches, focused in these areas (unit, m/s)

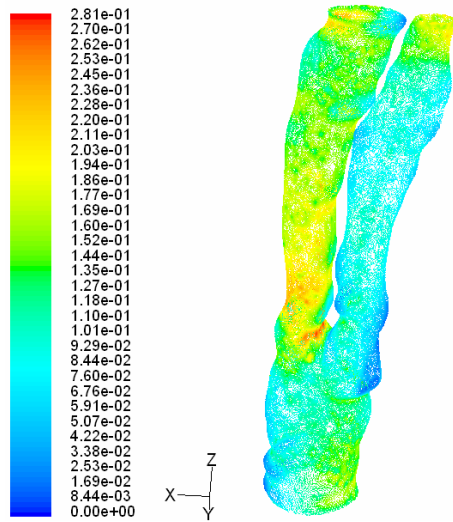

**Figure 6**. Wall shear stress (WSS) distribution for the diseased carotid artery.

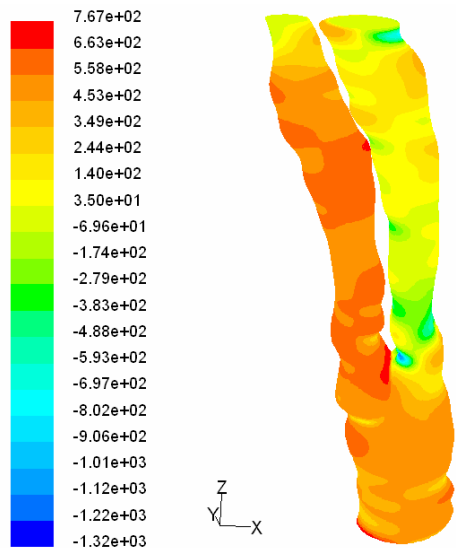

**Figure 7**. Pressure distribution for the diseased carotid artery (Pa)

**Conclusions:**

Computational analysis of the flow field inside the carotid arteries for a diseased male was performed using a commercial software. The flow was found similar to the flow inside a brancing channel. The geometry of the stenosis inside the diseased carotid artery defines the flow characterization. When the inner diameter gets smaller the flow rate will increase which will cause the high shear stress regions. The geometry due to the atheroschelerosis formation inside the internal and external branching caused the flow circulated and also low WSS. The upscaled experimental model of the diseased artery will be better to study to understand the flow regions with a higher spatial resolution in future. Also, the effect of the critical flow regions on the endothelial cells will be considered for the future studies to understand the mechanics of the real formation of the atherosclerosis.

**References**
[1] "Fry DL. Acute vascular endothelial changes associated with increased blood velocity gradients. Circ Res. 1968 Feb;22(2):165-97"

[2] "Caro CG, Fitz-Gerald JM, Schroter RC., Atheroma and arterial wall shear. Observation, correlation and proposal of a shear dependent mass transfer mechanism for atherogenesis. Proc R Soc Lond B Biol Sci. 1971 Feb 16;177(46):109-59"

[3] "Ku DN, Giddens DP, Zarins CK, Glagov S. Pulsatile flow and atherosclerosis in the human carotid bifurcation. Positive correlation between plaque location and low oscillating shear stress. Arteriosclerosis, 1985 May-June;(3):293-302"

[4] "Friedman SG, Kerner BA, Friedman MS, Moccio CG. Limb salvage in elderly patients. Is aggressive surgical therapy warranted? J Cardiovasc Surg (Torino). 1989 Sep-Oct;30(5):848-51"

[5] "Korner I, Blatz R, Wittig I, Pfeiffer D, Ruhlmann C. Serological evidence of Chlamydia pneumoniae lipopolysaccharide antibodies in atherosclerosis of various vascular regions regions. Vasa. 1999 Nov;28(4):259-63"

[6] "Glagov S, Rowley Da, Kohut Ri. , Atherosclerosis of human aorta and its coronary and renal arteries. A consideration of some hemodynamic factors which may be related to the marked differences in atherosclerotic involvement of the coronary and renal arteries. Arch Pathol. 1961 Nov;72:558-71"

[7] "DeBakey ME, Lawrie GM, Glaeser DH., Patterns of atherosclerosis and their surgical significance. Ann Surg. 1985 Feb;201(2):115-31"

[8] "Thubrikar MJ, Robicsek F., Pressure-induced arterial wall stress and atherosclerosis. Ann Thorac Surg. 1995 Jun;59(6):1594-603"

[9] "Lou Z, Yang WJ., A computer simulation of the blood flow at the aortic bifurcation with flexible walls. J Biomech Eng. 1993 Aug;115(3):306-15"

[10] "Perktold K, Rappitsch G., Computer simulation of local blood flow and vessel mechanics in a compliant carotid artery bifurcation model. J Biomech. 1995 Jul;28(7):845-56"

[11] "Cebral JR, Lohner R, Choyke PL, Yim PJ., Merging of intersecting triangulations for finite element modeling. J Biomech. 2001 Jun;34(6):815-9"